\documentclass[conference]{IEEEtran}

\usepackage{cite}

\usepackage{graphicx}

\ifCLASSINFOpdf
\else
\fi

\usepackage{algorithm}
\usepackage{algpseudocode}

\usepackage{array}

\usepackage[breaklinks=true]{hyperref}

\usepackage{comment}
\usepackage{array,booktabs}
\usepackage{multirow}
\usepackage{listings}
\usepackage{tabularx}

\usepackage{etoolbox}

\begin{document}

\title{Niji: Bitcoin Bridge Utilizing Payment Channels} 

\author{
\IEEEauthorblockN{Hiroki Watanabe, Shigenori Ohashi, Shigeru Fujimura,\\
Atsushi Nakadaira, Kota Hidaka}
\IEEEauthorblockA{NTT Service Evolution Labs., NTT Corp.\\
hiroki.watanabe.eh@hco.ntt.co.jp}
\and
\IEEEauthorblockN{Jay Kishigami\\}
\IEEEauthorblockA{Muroran Institute of Technology\\
jay@csse.muroran-it.ac.jp}
}
\makeatletter
\patchcmd{\@maketitle}
  {\addvspace{0.5\baselineskip}\egroup}
  {\addvspace{-1\baselineskip}\egroup}
  {}
  {}
\makeatother


\maketitle
\begin{abstract}
Bitcoin's enormous success has inspired the development of alternative blockchains, such as consortium chains. Several cross-chain protocols have been proposed as ways of connecting these universes of individual blockchains in a distributed and secure manner. In this paper, we present Niji, a new cross-chain protocol that allows parties to perform virtual Bitcoin payment securely on a consortium chain, without any trusted third-party or mediators. Our work focuses on the issue that it is difficult for a consortium chain's token to hold a stable market value, and Niji makes it possible for smart contract services to acquire means of payment in the consortium chain. With the Bitcoin payment channel built on the consortium chain, the process from payment to service provision runs autonomously without any interaction between parties. Niji introduces the concept of a transaction template to validate Bitcoin payments efficiently on different blockchains, and it allows a service provider to delegate all of its tasks for verifying state updates to a smart contract on the consortium chain. We also propose a novel bi-directional payment channel adapted for design of the Niji protocol, which can update payments non-interactively between parties. We implemented a prototype of the Niji protocol and conducted an experiment measuring the computational cost and latency that demonstrates the protocol’s feasibility on practical platforms.
\end{abstract}

\IEEEpeerreviewmaketitle

\section{Introduction}
Bitcoin is the most popular and successful cryptocurrency. Although Bitcoin has the largest market cap and holds a high volume of transactions per day, its growth as a currency has been gradual compared with its overheated growth as a speculative instrument. Bitcoin offers a lot of peer-to-peer payment opportunities, but its potential in most business use cases has not been fully realized. On the other hand, looking at the blockchain underlying Bitcoin, its technologies have attracted a great deal of attention from businesses, governments, and researchers.

The blockchain, a decentralized data management mechanism featuring tamper resistance, high availability, and data transparency, is expected to be a backbone of various commercial services besides those of the finance sector \cite{beyond}. Smart contracts, a new way to automate the execution of business work flows and their objective validation, has accelerated the evolution of blockchains into distributed application platforms. Smart contracts make blockchains suitable for a wide range of applications such as supply chains \cite{supply}, medical records \cite{medical}, online voting \cite{voting}, IoT platforms \cite{IoT}, transportation systems \cite{transport}, and energy trading \cite{energy}.

Indeed, while blockchain technology has already had a significant impact on society and business, there are many challenges still to be addressed. In particular, its industrial applications have limitations. Even though most enterprise applications require data privacy, transaction scalability, data reversibility, and protocol update-ability, these controls are not implemented on public blockchains such as Bitcoin and Ethereum.

A consortium blockchain (referred to as a ``consortium chain" after this), a type of blockchain whose consensus process is controlled by predetermined authorities, has been developed for tackling the challenges mentioned above. In particular, some groups of financial institutions and enterprises (e.g., Hyperledger Project, Ethereum Enterprise Alliance, and R3) have focused on developing implementations of blockchain platforms and modules for consortia. A number of companies have demonstrated proof-of-concept projects utilizing these platforms for enterprise use cases, to learn about the technology and its potential influence in their marketplaces. 

However, there is still a major challenge to overcome before consortium chains can reach mass market levels of penetration \cite{enter}. That is, ``payment" in a consortium chain is an opaque process, and it causes a problem affecting governance of the consortium. This is distinctly different from a public blockchain, which has a settlement function using cryptocurrency. Public blockchains are open for anyone to use and, thus, they gain some network effects by using common tokens for so many entities \cite{OnPub}. These tokens can be used directly for end-to-end payments between parties, and it is possible to exchange different assets efficiently. On the other hand, although consortium chains may also have tokens, in most cases, they hold no market value. One way to add valuable tokens to a consortium chain is to have external authorities (e.g., banks) ensure their value; in such case, final settlement is done via authorities outside the blockchain. The trouble is that the existence of a concentrated authority detracts from the advantages of blockchain technology such as transparency. Here, an ICO (Initial Coin Offering), a type of crowdfunding using cryptocurrency that many startups prefer, is an alternative way to make tokens valuable, but it is challenging for most businesses to make use of one because of legal complications and high costs associated with the process.

Another, more innovative and decentralized way is utilizing a cross-chain protocol to transfer the value of assets in the public blockchain to tokens of the consortium chain. There are several cross-chain solutions to address interoperability and scalability issues. The one type of cross-chain is to place trust in validator nodes such as Federated sidechain\cite{fed}, Cosmos\cite{cosmos}, and Polkadot\cite{polka}: the other is to have individual participants themselves watch for fraud on blockchains, like in AtomicSwap\cite{atomic} and Plasma\cite{Plasma}. In any case, the common premise is that the value of assets in the main chain are transferred to tokens in another chain.

This paper presents Niji, a novel cross-chain protocol that supports atomic and secure payment on a consortium chain without any authorities. The key difference from previous cross-chain solutions is that our protocol does not transfer value to the consortium chain. Instead, by utilizing a payment channel, the protocol enables virtual Bitcoin payments on the consortium chain by verifying payments with smart contracts. Our solution is simple, yet can be easily introduced, and it is practical in the context of the consortium. Since the protocols of each chain are logically decoupled, there is no need to maintain a large and complex system of cooperation, which leads to significant cost savings. The consortium can focus on service operations rather than economic issues. In addition, from the perspective of Layer 2, our contribution scales opportunities to use bitcoins by enabling execution of secure off-chain Bitcoin payments on diverse blockchains. This is different from the approach of forming one network like Lightning, but it will be possible to use Bitcoin payment in the various universes of individual blockchains.

\subsubsection*{Outline}
Section II introduces the elements necessary to understand the rest of this article. Section III contains a detailed description of the basic Niji protocol using a uni-directional payment channel. The extension to a bi-directional channel is shown in section IV. Section V analyzes the security of the Niji protocol and its feasibility through an experimental implementation. We compare Niji with other cross-chain protocols and discuss possible extensions of this protocol in section VI and end this paper in section VII.

\section{Building Blocks}
This section establishes the necessary building blocks for understanding the Niji protocol. In the following, we introduce payment channels, which are techniques for Bitcoin off-chain payment; then we describe the Ethereum Virtual Machine with which the arbitrary code of smart contracts is executed.

\subsection{Payment Channels}
The payment channels achieve end-to-end secure payment for off-blockchain trading. Niji utilizes a payment channel as a sub-protocol on the cross-chain protocol. The simple payment channel that was first discussed by Hearn and Spilman \cite{simp} allows a payer to send a payee numerous payments without committing all of the transactions to the blockchain. The channel is essentially uni-directional; that is, the payer can send bitcoins to the payee, but the payee cannot send in the opposite direction. In a simple payment channel, only two transactions are stored in the blockchain: a {\it funding transaction} and a {\it settlement transaction}. The funding transaction is used to open the channel. The transaction deposits a payer's bitcoins into a 2-of-2 multi-signature account managed by the payer and payee. Conversely, the settlement transaction is used to close the channel. It eventually performs the transfer of funds on the blockchain and settles the balances of the payer and the payee.

While the channel is open, the payer can update his/her payment multiple times within the range of funds deposited in the multi-signature account. For example, in a first-time payment, the payer creates and signs a transaction that transfers 0.1 BTC to the payee from 1 BTC of the multi-signature account. Next, in a second payment to send the remaining 0.4 BTC, the payer creates another transaction that sends 0.5 BTC to the payee and updates the state of the channels. The payer cannot broadcast any of those transactions to the Bitcoin network, since these transactions do not have the required signature of the payee. In this paper, we refer to such an incomplete transaction as an {\it update transaction}, which specifies a funding transaction as input and includes only the signature of the payer. To transfer funds from the multi-signature account, the transaction requires signatures of both the payer and payee. Therefore, only the payee has the right to sign and broadcast the last state of the channel (i.e., settlement transaction with 0.5 BTC) at an arbitrary timing. To protect the payer from the risk that the payee does not respond and does not cooperate by broadcasting any state of the channel, a time-lock that refunds the whole 1.0 BTC to the payer is applied to the output's script of the funding transaction. There are two different types of time-lock; {\it CheckLockTimeVerify} (CLTV), which is an opcode specified in BIP65\cite{bip65}, allows users to create a transaction whose outputs are available until a concrete time in the future. On the other hand, {\it CheckSequenceVerify} (CSV), introduced in BIP112\cite {bip112}, specifies a relative time. When a transaction output including OP\_CSV is stored in the blockchain, it is necessary to wait for the specified block confirmations until the transaction is spendable again.

The simple payment channel described above is substantially uni-directional; that is, the payer can send bitcoins to the payee, but not in the opposite direction. In a bi-directional channel, the flow can go both ways. There are two well-known proposals for Bitcoin bi-directional channels. The first is called Duplex Micropayment Channels, proposed by Decker and Wattenhofer \cite{dmc}. It achieves bi-directional payment channels by using two uni-directional payment channels with a finite lifetime. The second is Lightning Network by Poon and Dryja \cite{ln}, which allows the channel to remain open indefinitely, relying on punishments to promote honesty among parties.

Moreover, regarding the existing solutions using payment channels, there is an interactive process from payment to service provision. For each payment, the payee must confirm whether the channel state is valid. When receiving an update transaction through an off-chain network, the payee must validate the transaction format and verify the included signature. Then, the payee can provide a service, say a WiFi hotspot service. If the payment of the update transaction is invalid, the provision will obviously be refused. Our solution, Niji, can automate this verification process by using smart contracts on the consortium chain.

\subsection{Ethereum Virtual Machine}
We assume that Ethereum Virtual Machine (EVM) is the execution environment of consortium chain smart contract in the Niji protocol implementation. EVM executes ordinary transactions and treats smart contract bytecode as a special transaction. Specifically, each smart contract is given its own storage to keep the state of the contract. Our protocol can run in other environments; however, we should emphasize the utility of EVM from the following aspects. First, EVM has been ported to several private blockchain platforms. The blockchains released by various companies, open source communities, universities, etc., currently have different technologies and frameworks, and market fragmentation is occurring as a result. EVM was originally intended as a runtime environment for smart contracts on the Ethereum platform, and it has since been integrated into several blockchain platforms for enterprise, such as Quorum, Monax, Hyperledger Burrow, and Ethereum on Azure. In addition, go-ethereum, an official Ethereum client of the Go implementation, can build a consortium blockchain using a proof-of-authority algorithm (EIP225\cite{clique}). These facts mean that EVM-based smart contracts can run compatibly on multiple platforms, as described above.

As a second reason, EVM smart contracts are less flexible than other Turing-complete smart contracts like Hyperledger Fabric chaincode. There are trade-offs between the flexibility that a platform provides and the security of the code which developers write for it. EVM's target is basically to provide a ``public" environment where arbitrary code of smart contracts and other operations can be executed on the Ethereum. Therefore, EVM inevitably has more restrictions on executions of smart contracts than does a platform that is secure only for closed environments. For example, the ``gas system" is a restriction to prevent eternal recursion and cycles, and it encourages developers to write efficient code (the gas system is described in section IV-A). Therefore, a protocol designed to work only on EVM may also be able to run on other Turing-complete platforms like Hyperledger Fabric.

\section{Niji Protocol}
\begin{figure*}[ht]

\centering
\includegraphics[width=\textwidth]{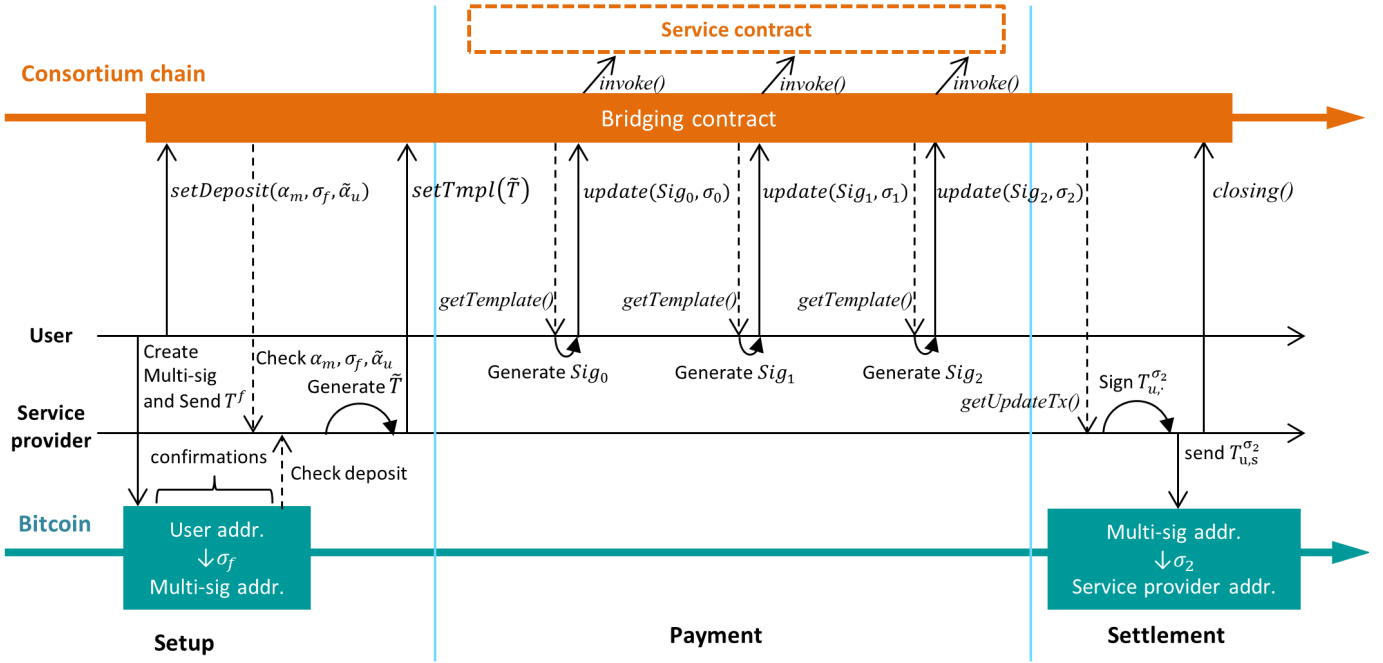}
\caption{Overview of the Niji Protocol. Starting on the left, a user commits to a payment to the bridging contract three times ($i=0,1,2$). Straight solid lines with arrows represent sending transactions to the blockchain networks. Straight dashed lines with arrows represent retrieving data from the blockchains.}
\label{fig_sim}
\vspace{-0.8em}
\end{figure*}

The Niji protocol is essentially based on a Bitcoin payment channel. As described in section II, payment channels come in two types: uni-directional and bi-directional. For an intuitive understanding, we will first explain the uni-directional channel-based protocol (the basic protocol) and then how it is extended to the more practical bi-directional protocol.

\subsection{Overview}
Niji allows Bitcoin payments within the payment channel to trigger automatic execution of the service deployed on the consortium chain. To achieve this, the payee does not locally verify an update transaction received from the payer; instead, the EVM smart contract validates these in a consortium chain. Niji provides an autonomous process from payment to execution of a contract, without any mediators. 

To begin with, let us identify the roles representing different functionalities in the Niji system:
\begin{itemize}[\IEEEsetlabelwidth{}]
\item[a)] {\it Networks}: Each node of the Niji system has connections to two blockchain networks: the Bitcoin blockchain and a consortium blockchain. Unlike Bitcoin, the consortium chain network restricts unauthorized access to the network for secure transactions. In this network, nodes do not need to be block generators (i.e., a member of a consortium authority, like a bitcoin miner), but they need to have the right to read and write data (i.e. issue transactions) to the blockchain.
\item[b)] {\it Nodes}: The {\it user} and {\it service provider} represent the parties in this protocol. The user is a payer node which pays bitcoins to the service provider for usage of the services operated on the consortium chain. The user makes Bitcoin payments which trigger execution of a smart contract on the consortium chain while the payment channel is open. The last state of payments is broadcasted as a settlement transaction by the service provider and is eventually stored in the blockchain. The service provider is a payee node with the intention to earn money by providing some kind of service (e.g., sharing a resource like energy for microgrids\cite{energy}) on the consortium chain. The service provider creates a service contract for providing the service and shares it via the consortium chain. It receives a Bitcoin payment from the user as the usage fee of this contract via the Niji protocol.
\item[c)] {\it Smart Contracts}: Two smart contracts are deployed on the consortium blockchain. The {\it bridging contract} has functions to manage the user's payments instead of the service provider doing so. It receives payments and validates them; then, it invokes the service contract's function. The {\it service contract} has an interface to invoke the core functionality of the service, and the interface is opened for the bridging contract. The two contracts are created and deployed on the consortium blockchain before beginning the Niji protocol.
\end{itemize}

\begin{table}[tbp]
\renewcommand{\arraystretch}{1.0}
\caption{Structure of Transaction Template}
\label{timing}
\centering
\begin{tabular}{llll}
 \toprule Field &  & Value (example)\\
 \midrule
 Version &  & 02000000 \\
 Input count &  & 01 \\
 \cmidrule(l){1-2}\cmidrule(lr){3-3}
 Input {[}0{]} & Previous output & \textless{}funding tx hash\textgreater{} \\
  & Index & \textless{}index of previous output\textgreater{} \\
  & Script & only \textless{}redeemScript of previous  \\
  & & output\textgreater{} {\bf (no signatures)} \\
  & Sequence &  FFFFFFFF \\
 \cmidrule(l){1-2}\cmidrule(lr){3-3}
 Output count &  & 02 \\
 \cmidrule(l){1-2}\cmidrule(lr){3-3}
 Output{[}0{]} & Value & {\bf nil} \\
  & Script & OP\_DUP OP\_HASH160 \\
  & &\textless{}service provider's pubkey hash\textgreater{}\\
  & & OP\_EQUALVERIFY OP\_CHECKSIG \\
 Output{[}1{]} & Value & {\bf nil} \\
 (change given & Script & OP\_DUP OP\_HASH160 \\
 back to user)& & \textless{}user's pubkey hash\textgreater{} \\
 & & OP\_EQUALVERIFY OP\_CHECKSIG \\
 \cmidrule(l){1-2}\cmidrule(lr){3-3} 
 Locktime &  & 00000000 \\
 Hash Type &  & 01000000 \\
 \bottomrule
\end{tabular}
\end{table}

The protocol itself is designed as follows. The core idea is to automate the process of updating a payment non-interactively. This is achieved using incomplete Bitcoin transactions, which we name {\it transaction templates}, where information on the remittance amount and signatures are missing. The service provider registers a transaction template to the bridging contract during the setup phase and publishes it on the consortium chain in advance. Table I shows an example of a transaction template. For simplicity, the example uses a non-Segwit (Segregated witness) transaction. (Segwit version is described in Appendix A.)

When making a payment, a user creates a signature corresponding to the payment referring to the transaction template and submits only the signature and remittance amount to the bridging contract. The bridging contract verifies the provided signature on behalf of the service provider and validates the payment by using the registered transaction template, user's signature, and remittance amount. If valid, the payment is automatically approved, which means that the service provider accepts the update transaction in the simple payment channel.

\subsection{Protocol details}
Figure 1 shows the three major phases of the Niji protocol from setup to settlement. We describe each phase in the following.

\subsubsection{Setup}
The setup involves opening the Bitcoin simple payment channel and registering a transaction template in the consortium chain. First, to open the typical simple payment channel shown in section II-A, the user and service provider create a 2-of-2 multi-signature address $\alpha_{m}$ and broadcast a funding transaction $T^{f}$ to the Bitcoin network. This transaction $T^{f}$ sends the user's bitcoins to the multi-signature accounts, where the amount is denoted as $\sigma^{f}$. For a refund on $T^{f}$, we use a relative time-lock by applying OP\_CSV \cite{bip65} to the multi-signature accounts; if the time-lock expires, the user can securely return all deposits to his/her wallet. Then, the user broadcasts a transaction ${}^{C}T(setDeposit(\alpha_{m}, \sigma^{f}, \tilde \alpha_{u}))$ as proof of opening the channel, where ${}^{C}T(func)$ is a consortium chain transaction calling a function of the bridging contact. The function $setDeposit(\alpha_{m}, \sigma^{f}, \tilde \alpha_{u})$ stores deposit information including the multi-signature address $\alpha_{m}$, deposit amount $\sigma^{f}$ in the consortium chain, and an address $\tilde \alpha_{u}$. Here, $\tilde \alpha_{u}$ is a special Ethereum-style address which is derived from the Bitcoin public key. How $\tilde \alpha_{u}$ is used for verification is explained in the next subsection.

Next, after making enough confirmations (e.g., six confirmations), the service provider confirms that $T^{f}$ is stored in the Bitcoin blockchain and generates a transaction template $\tilde T$, which is a deformed transaction lacking signatures and the remittance amount. $\tilde T$ specifies $T^{f}$ as input and corresponds to the complete transaction $T^{\sigma}_{us}$, which represents a valid Bitcoin transaction sending $\sigma$ bitcoins from the multi-signature address to the service provider's address. The subscripts $u$ and $s$ represent that $T^{\sigma}_{us}$ has the signatures of the user $u$ and the service provider $s$. $T^{\sigma}_{us}$ is a 5-tuple including $\tilde T$ as follows:

\begin{equation}
T^{\sigma}_{us}=(\tilde T,Sig^{u},Sig^{s},\sigma,\sigma^{c})
\end{equation}
where $Sig^{u}$ and $Sig^{s}$ are the respective signatures of $u$ and $s$, and $\sigma^{c}$ is amount of change returned to $u$.

At the end of setup, the service provider broadcasts a transaction ${}^{C}T(setTmpl(\tilde T))$ to the consortium chain network. A function $setTmpl(\tilde T)$ stores $\tilde T$ in the storage space of the bridging contract on the consortium blockchain.

\subsubsection{Payment}
This phase allows the bridging contract to validate each payment on behalf of the service provider. The bridging contract has a function $getTemplate()$ for retrieving the stored $\tilde T$ and a function $update (Sig, \sigma)$ for updating the signature value and the remittance amount. First, the user obtains $\tilde T$ from the bridging contract by using $getTemplate()$ and validates it if its format is correct. Then, when making a payment, the user provides the user's signature $Sig^{u}$ and $\sigma$ for the bridging contract as a proof of payment using $update (Sig, \sigma)$. The user generates $Sig^{u}$ as follows:

\begin{equation}
 modtx　= SignatureForm(\tilde T, \sigma)
\end{equation}
\begin{equation}
 Sig^{u} = EcdsaSign(sk^{u}, Sha256d(modtx))
\end{equation}

$ modtx $ is the modified transaction form just before it is signed, which is created by $\tilde T$ and $\sigma$, where the redeemScript of $T^{f}$ is placed into $ modtx $'s input and a hash type constant is temporarily appended to the end. To generate signatures, Bitcoin uses ECDSA (the Elliptic Curve Digital Signature Algorithm) over the standard elliptic curve secp256k1, which requires possession of the signing secret key $sk^{u}$ and a sha256 double hash of $modtx$ to be signed \cite{hardB}. ${}^{C}T(update(Sig, \sigma))$ is broadcasted to the consortium blockchain network by the user and stored in the blockchain, and the bridging contract validates its payment automatically. The validation includes confirming whether the value of $\sigma$ is under the value of $\sigma^{f}$ and verifying that the signature $Sig^{u}$ is correct. Algorithm 1 lists the pseudo-code of the payment updating process in the bridging contract.

\begin{algorithm}[tbp]
\caption{Updating a payment}
\label{array-sum}
\begin{algorithmic}[1]
    \Procedure {UpdatePayment}{$Sig,\sigma$}
    \State Obtain $\tilde T$, $\tilde \alpha_{u}$, $\sigma^{f}$ from contract storage;
    \State $fee \gets$ configured bitcoin transaction fee
    \State $S_{con} \gets$ deployed service contract
    \State $\sigma_{l} \gets 0$ or latest remittance amount
    \If{$\sigma_{l}<\sigma$ and $\sigma\leq(\sigma^{f}-fee)$}
        \State $modtx \gets SignatureForm(\tilde T, \sigma)$
        \State $h \gets  Sha256d(modtx)$
        \State $(v,r,s) \gets Sig$ 
        \State $pk \gets EcdsaRecover(h, v, r, s)$
        \State $addr \gets EthereumAddress(pk)$\\ \Comment {Convert  Bitcoin's ECDSA public key to Ethereum-style Address} 
        \If{$\tilde \alpha_{u} = addr$}
            \State $s \gets true$
            \State $result \gets S_{con}.invoke()$
            \State Save ($Sig$, $\sigma$) in contract storage;
            \State \textbf{return} $result$
        \EndIf
\EndIf
\State \textbf{return} $fail$
\EndProcedure
\end{algorithmic}
\end{algorithm}

\begin{algorithm}[tbp]
\caption{Generating an update transaction}
\label{array-sum}
\begin{algorithmic}[1]
    \Procedure {GetUpdateTx}{$ $}
    \State Obtain $\tilde T$, $\sigma_{i=k}$, $Sig^{s}_{i=k}$ from contract storage;\\ \Comment {$k$ is last increment number;} 
    \State $fee \gets$ configured bitcoin transaction fee
    \State $\sigma^{c}\gets\sigma^{f}-\sigma_{k}-fee$
    \State $T^{\sigma}_{u\cdot}=(\tilde T,Sig^{u},\cdot,\sigma,\sigma^{c})$
\State \textbf{return} $T^{\sigma}_{u\cdot}$
\EndProcedure
\end{algorithmic}
\end{algorithm}

\subsubsection*{Verification}
Verification of $Sig^{u}$ in the EVM implementation requires a little ingenuity. Since signature verification using the native opcodes of EVM has too much overhead, we use ``precompiled contracts" \cite{pcon}, which allow complex cryptographic computations to be used in EVM. A precompiled contract is a specific pre-defined optimized function that runs outside the EVM, but it can be called from a normal smart contract that runs inside the EVM. $EcdsaRecover$, one such precompiled contract, can verify an elliptic-curve signature and recover its public key $pk$ as follows:

\begin{equation}
EcdsaRecover(h, v, r, s) = pk
\end{equation}
where $h$ is a 32-byte message to be signed, and $(v, r, s)$ is the ECDSA signature for the message (v is the recovery id, a 1 byte value specifying the sign and finiteness of the curve point \cite{pcon}). This signature is compatible with a Bitcoin-style signature which has a strict DER (Distinguished Encoding Rules) format \cite{bip66}; therefore, the public key $pk$ is recovered using $h=Sha256d(modtx)$ and $(v, r, s)^{u}=Sig^{u}$ through $EcdsaRecover$. Finally, the verification is successful if the Ethereum-style address converted from $pk$ matches $\tilde \alpha_{u}$ stored in the bridging contract. Accordingly, $Sig^{u}$ can be verified using $\tilde T$, $\sigma$, and $\alpha_{u}$, which are registered in the bridging contract. Here, the reason for the verification being done via an Ethereum-style address is the specification of Solidity, a high-level language for implementing Ethereum smart contracts. (The verification process in Solidity is detailed in Appendix B.)

\subsubsection*{Update payments} 
The total payment amount is updatable multiple times up to the amount of $\sigma^{f}$ that the user first deposited. We denote $\sigma$ and $Sig^{u}$ after update $i$ as $\sigma_{i}$ and $Sig^{u}_{i}$ for $i = 0,\ldots,n$. The user adds a new payment to the current payment amount $\sigma_{i}$ and replaces $\sigma_{i}$ and $Sig^{u}_{i}$ with $\sigma_{i+1}$ and $Sig^{u}_{i+1}$ by broadcasting ${}^{C}T(update(Sig^{u}_{i+1}, \sigma_{i+1}))$. $\sigma_{i}$ must be under $\sigma_{i+1}$ ($\sigma_{i}<\sigma_{i+1}$), since the payment channel is uni-directional. This is checked in the bridging contract.

\subsubsection*{Triggering contract's function}
$invoke()$, a function of the bridging contract, can call services provided by the service contract. After validation of a payment, the bridging contract executes $invoke()$, triggering the service contract's function corresponding to the payment. This process is an internal communication between smart contracts that protects nodes in the procedure. Thus, the procedure of transaction ${}^{C}T(update(Sig, \sigma))$ runs from payment to provision safely and automatically.

\subsubsection{Settlement}
In the settlement phase, the service provider obtains an update transaction from the bridging contract, signs it, and broadcasts it to the Bitcoin network as a settlement transaction. Algorithm 2 lists the pseudo-code of the bridging contract. $getUpdateTx()$, a function of the bridging contract, returns the update transaction $T^{\sigma}_{u\cdot}$ (the service provider's signature is replaced with $\cdot$, which represents “not yet signed”) to the service provider. After signing $T^{\sigma}_{u\cdot}$ with its own secret key, the service provider broadcasts the complete transaction $T^{\sigma}_{us}$, i.e., the settlement transaction for Bitcoin. Finally, the service provider broadcasts a transaction ${}^{C}T(closing())$ that makes any further payments to the bridging contract unacceptable.

\section{Niji with bi-directional channel}
The previous section described the basic specifications of the Niji protocol using the simple payment channel. As mentioned there, the simple payment channel cannot send in the opposite direction, which means that cancellation of payments is not allowed. Security in the simple payment channels is based on the fact that the recipient of payment does not have an incentive to broadcast the old state of the channel; that is, the latest update transaction always brings the maximum benefit to the recipient. However, if a payment in the opposite direction occurs, the recipient has an incentive to broadcast the old state of the channel at any point in time. In the Niji protocol, the service provider can broadcast and settle an old update transaction any time, in spite that the parties have agreed to cancel the payment.

Hence, to achieve secure cancellation, we considered using Duplex Micropayment Channels (DMC) or Lightning Network channels, both bi-directional channels, in the Niji protocol. In DMC, since both parties create two uni-directional channels, the service provider, as well as users, needs to deposit funds. The drawback of using this design is if the service provider has numerous customers, a large amount of money will be tied up. On the other hand, the Lightning Network payment channel requires both parties to exchange their signatures each time a channel state is updated. This interactive process hinders automation from payment to execution of a contract, one of the advantages of Niji. Also, both parties have to monitor not only the consortium chain but also the Bitcoin network to prevent counterparty fraud.

To avoid the above problems, we propose a new non-interactive channel design suitable for the Niji protocol. In particular, to ensure autonomous blockchain cooperation, the bi-directional channel is designed to meet the following requirements.
\begin{itemize}[\IEEEsetlabelwidth{Req.}]
\item[Req. 1.] The user can update their payment without interacting with the service provider node.
\item[Req. 2.] The user is not required to monitor the Bitcoin network and does not have to be on-blockchain at all times.
\end{itemize}
Note that unlike other bi-directional channels, Niji's bi-directional channel takes into account the following special consideration: almost all payments are from the user to the service provider; transactions in the opposite direction occur only a few times (such as cancellation of payment). The central idea of our channel is to introduce this asymmetry in the procedures in each direction. Parties can update normal payments non-interactively, but temporarily collaborate when canceling payments. Our channel is not completely bi-directional like Lightning, but it is adequate for most of the targeted use cases. In following subsections, we describe the separate channel designs for normal payment and cancellation.

\begin{figure}[tbp]
\centering
\includegraphics[trim=0 0 0 0,clip, width=0.36\textwidth]{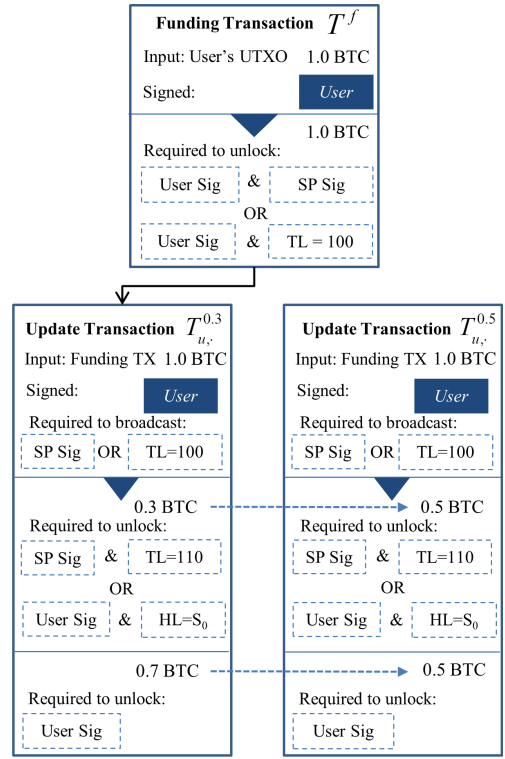}
\caption{Examples of update transactions in our bi-directional channel between user and service provider (SP). TL and HL are parameters for unlocking the time-lock and hash-lock, respectively.}
\label{fig_sim}
\end{figure}

\subsection{Payments in the normal direction}
Our channel consists of three type of transaction: funding, update, and settlement. In the normal direction of payment, the channel acts like a simple payment channel. That is, in each payment, the user increases his/her remittance amount and never presents a remittance amount lower than the previous amount. The update transactions have two outputs: the first one specifies the remittance amount, and the second specifies the amount of change given back to the user. The spending output for the remitted amount is restricted by the following additional conditions.

\subsubsection*{Condition 1}
The spending of the output is restricted by a time-lock. The service provider can unlock the script with its signature only after a specified time-lock expires.

\subsubsection*{Condition 2}
The spending of the output is restricted by a hash-lock. The user can unlock the script with the hash's pre-image and its signature.

These conditions are expressed using a Bitcoin script, and either condition is fulfilled by a script in the inputs of the next transaction. Figure 2 shows examples of update transactions that include these conditions in the first output. The examples indicate that the user creates a new update transaction $T^{0.5}_{u\cdot}$ with the remittance amount increased from {0.3BTC} to {0.5BTC}.

The service provider can move the gained remittance to its own wallet by using condition 1 in the first output. However, it is impossible for the service provider to immediately fulfill the condition as it is encumbered with the time-lock. It can obtain the funds only after the time-lock expires. In the example in Figure 2, $TL$ is denoted as the time until the transaction becomes spendable again; if $TL$ = 110, it takes 110 blocks from the point where the transaction occurred. The time-lock gives the user a period for preventing any dishonesty on the part of the service provider.

Condition 2 exists as a user's countermeasure against betrayal by the service provider. The hash-lock requires the subsequent input to include the corresponding pre-image (i.e., $S_{0}$ in Figure 2) of the hash in order to be spendable. A hash is passed to the user beforehand by the service provider, and if the user obtains the pre-image of the hash, he/she can spend the update transaction.

Now, with respect to the hash-lock, it should be noted that the same pre-image $S_{0}$ is reused before and after the update, as shown in Figure 2. This is in stark contrast to the Lightning payment channel, where the pre-images are recreated each time a state is updated. In Lightning, both parties need to exchange information including old pre-images and the hash values of new pre-images to create the next update transaction; thus they are required to interact with one another and to be always online in the meantime. In the proposed channel in Niji, it is not necessary to recreate new images each time the state is updated. This satisfies the first requirement (Req. 1), where interactions with parties are unnecessary, and it enables the service provider to delegate its update processing to the smart contract.

Additionally, the deadline of the time-lock $TL_{u}$ of an update transaction is always set to be after the time-lock $TL_{f}$ of a funding transaction. For example, if $TL_{f}$ is set to 100 blocks, $TL_{u}$ must be set to 101 blocks or later. These time-lock settings definitely guarantee that the service provider can not move the gained funds until the channel expires. Therefore, the user does not have to monitor the Bitcoin network continuously while the channel is alive in order to be sure that the service provider has not committed fraud; therefore, these restrictions result in the fulfillment of the second requirement (Req. 2).

Now, let us describe how the above bi-directional channel for normal payments is applied to the Niji protocol. In particular, it is achieved by adding constraints to the output of the transaction template which is part of an update transaction. Table II shows an example of a whole transaction template and its output[0]'s redeemScript including condition 1 and condition 2 in Figure 3. We will express the transaction template $\tilde T$ as $\tilde T(TL, HL)$; therefore, the template having the redeemScript shown in Figure 3 can be represented as $\tilde T(110, S_{0})$. The procedure for a normal payment is no different from the basic Niji protocol described in section III, except that the redeemScript must be shared with the user; hence, we modify the $setTmpl$ function from $setTmpl(\tilde T)$ to $setTmpl(\tilde T(TL, HL), redeem)$, where $redeem$ is a redeemScript corresponding to $\tilde T(TL, HL)$. Of course, the bridging contract can verify the correctness of $redeem$, which includes whether $redeem$ is the same as the hash of output[0] script in $\tilde T(TL, HL)$, whether the $TL$ included in $redeem$ is longer than the number of blocks specified in $TL_{f}$ of a funding transaction, and so on.

\begin{table}[tbp]

\renewcommand{\arraystretch}{1.0}

\caption{Structure of Transaction Template in Bi-directional Channel}
\label{timing}
\centering

\begin{tabular}{llll}
 \toprule Field &  & Value \\
 \midrule
 Version &  & 02000000 \\
 Input count &  & 01 \\
 \cmidrule(l){1-2}\cmidrule(lr){3-3}
 Input {[}0{]} & Previous output & \textless{}funding tx
 hash\textgreater{} \\
  & Index & \textless{}index of previous output\textgreater{} \\
  & Script & only \textless{}redeemScript of previous  \\
  & & output\textgreater{} {\bf (no signatures)} \\
  & Sequence &  FFFFFFFF \\
 \cmidrule(l){1-2}\cmidrule(lr){3-3}
 Output count &  & 02 \\
 \cmidrule(l){1-2}\cmidrule(lr){3-3}
 Output{[}0{]} & Value & {\bf nil} \\
  & Script & OP\_HASH160 \\
  & & {\bf \textless{}hash of redeemScript in Fig. 3\textgreater{} }\\
  & & OP\_EQUAL \\
 Output{[}1{]} & Value & {\bf nil} \\
 (change given & Script & OP\_DUP OP\_HASH160 \\
 back to user)& & \textless{}user's pubkey hash\textgreater{} \\
 & & OP\_EQUALVERIFY OP\_CHECKSIG \\
 \cmidrule(l){1-2}\cmidrule(lr){3-3} 
 Locktime &  & 00000000 \\
 Hash type &  & 01000000 \\
 \bottomrule
\end{tabular}
\end{table}

\begin{figure}[tbp]
\renewcommand{\arraystretch}{1.0}
\footnotesize
\begin{tabularx}{\columnwidth}{ll}
 \midrule
  & OP\_IF \\
  & \qquad  OP\_HASH160 \\
  & \qquad  \textless{}hash of $S_0$\textgreater{} \\
  & \qquad   OP\_EQUALVERIFY \\
  & \qquad   \textless{}user's pubkey\textgreater{} \\
  &  OP\_ELSE \\
  & \qquad   110 OP\_CSV\\
  & \qquad   OP\_DROP \\
  & \qquad   \textless{}service provider's pubkey\textgreater{} \\
  & OP\_ENDIF \\
  & OP\_CHECKSIG \\
 \midrule
\end{tabularx}
\caption{Example of redeemScript in output[0]}
\end{figure}

\subsection{Cancellation in the opposite direction}
Next, we describe how the protocol behaves in the opposite direction, i.e., when payments are canceled. Unlike a normal payment, several agreement operations between parties are required to cancel payments. The user and service provider need to interact with one another and exchange materials for agreement on the cancellation. The key steps are as follows.
\begin{itemize}[\IEEEsetlabelwidth{STEP}]
\item [STEP1.] The user sends a cancel request to the service provider.
\item [STEP2.] The service provider creates a new pre-image and returns the hash value of it.
\item [STEP3.] The user sets the received hash and submits a signature for the new update transaction.
\item [STEP4.] The service provider validates the user's signature and discloses the old pre-image of previous update transaction.
\end{itemize}

\begin{figure}[tbp]
\centering
\includegraphics[width=0.36\textwidth]{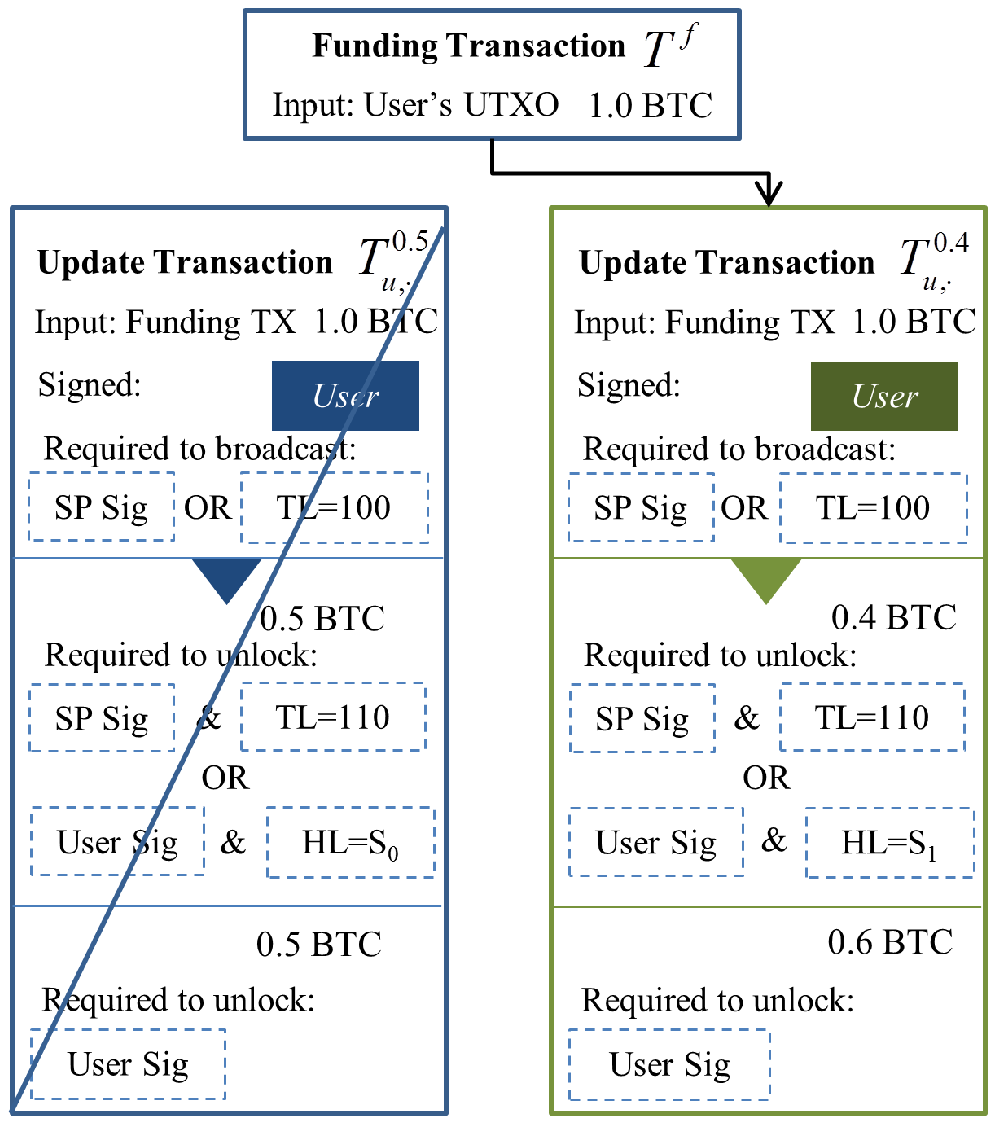}
\caption{Example of update transactions for cancellation of payment in Niji bi-directional channel. Even if the service provider broadcasts an old update transaction, the user can get back the funds by using the disclosed pre-image $S_{0}$.}
\label{fig_sim}
\end{figure}
\begin{figure}[tbp]
\centering
\includegraphics[width=0.5\textwidth]{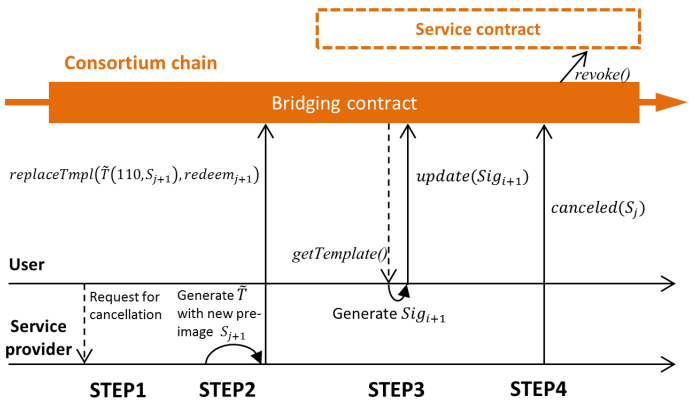}
\caption{Process for cancellation of payment in Niji protocol}
\label{fig_sim}
\end{figure}
With the completion of the above four steps, the cancellation is deemed to be agreed upon by the parties. Afterwards, the service provider can no longer broadcast the old update transaction. Figure 4 shows that the old update transaction is replaced with a new update transaction. The remittance amount is reduced from {0.5BTC} to {0.4BTC} with a new pre-image $S_{1}$. If the service provider broadcasts an old update transaction, the disclosed pre-image $S_{0}$ allows the user to get back all funds. (Techniques similar to this can be found in HTLCs or Lightning Network.) Moreover, if the procedure breaks down or terminates before the agreement, it can be regarded that an agreement has not been reached. For example, if the cancellation request is invalid, the service provider can not respond to the user's request.

The cancellation of payment operations is illustrated in Figure 5. Here, it is paramount that all materials required for cancellation are exchanged via the consortium chain. This is achieved by registering a new transaction template which includes the hash value of a new pre-image on the consortium chain (equivalent to STEP 2). $replaceTmpl(\tilde T(110, S_{j+1}), redeem_{j+1})$ replaces the existing $T(110, S_{j})$ and $redeem_{j}$ with a new $T(110, S_{j+1})$ and $redeem_{j+1}$ in the storage space of the bridging contract, where the increment count $j$ represents the number of cancellations. Then, the user makes $Sig_{i+1}$ with the new amount, which is the reduced remittance after cancellation, and registers $Sig_{i+1}$ to the bridging contract via the $update$ function (equivalent to STEP 3). Finally, the service provider discloses $S_{j}$ via the $canceled(S_{j})$, which publishes the pre-image of the previous update transaction on the consortium chain (equivalent to STEP 4).

For secure trading, these procedures can be strictly validated by using a smart contract on the consortium chain, i.e., a bridging contract, which prevents service providers from broadcasting old update transactions. In addition, success of cancellation in the contract can automatically trigger the service contract's functions such as $revoke()$, which reverts the state back to one before the payment was canceled.

\section{Analysis}
The new bi-directional payment channel for the Niji protocol enables autonomous cooperation between blockchains. In the following, we analyze the assumptions of security and evaluate the feasibility of our prototype.

\subsection{Security}
The security of the Niji protocol essentially depends on the safety of the payment channel. Although different channels may be applied to Niji in the future (see section VI-C), we will consider the safety of only the bi-directional channel proposed in this paper. In Niji, since the service provider always broadcasts a settlement transaction, it has more discretion when it comes to making the settlement in comparison with the user. Therefore, the major security risk we consider is an act of betrayal of the service provider. The service provider could potentially behave dishonestly in three ways.
\begin{itemize}[\IEEEsetlabelwidth{}]
\item [1.] No settlement is made at all.
\item [2.] Despite that a cancellation was approved, the service provider broadcasts an old update transaction.
\item [3.] The contract deployed by the service provider does not run properly.
\end{itemize}
In the first case, the service provider abandons the protocol itself and refuses to broadcast a settlement transaction within the channel expiration. To prevent such a deadlock of funds, the output script of a funding transaction allows the user to return all deposits to his/her wallet after the funding transaction's time-lock expires; that is, the user can broadcast the settlement transaction with only the user's signature. The second case is that the service provider does not broadcast the latest channel state that reflects cancellation and instead broadcasts an old channel state which has more incentives. This fraud can be prevented by using the disclosed pre-image at the end of the cancellation procedure, as described in the previous section. With the user's signature and disclosed pre-image, the user can spend the output of a settlement transaction under condition 2 described in section IV-A.

The third case is caused by an issue with the contract itself rather than the operation of the payment channel. All contracts in the Niji protocol, which include Bitcoin scripts and smart contracts on the consortium chain, are provided by the service provider. The user must confirm that the contract is the expected one to ensure safety. Unlike conventional cross-chains using an oracle outside of the blockchain, all of our contracts are on a consortium chain and the entire transaction history is shared among the participants sharing the blockchain; therefore, it is possible for the user to re-validate the state changes and observe whether the contract works properly via the blockchain. In addition, as required by the security level, it is assumed that a healthy consortium chain network continues; hence in the consortium chain network, the block creators are required to be honest and faithful to the protocol; e.g., they are encouraged to use a consensus algorithm with transaction finality like PoA and PBFT.

Incidentally, Plasma \cite{Plasma}, a cross-chain framework that has similar aspects to our approach, has precautions that involve cooperation between blockchains; e.g., a child blockchain is enforced to roll back by submitting proof of fraud to the parent blockchain. In our approach, there is no close cooperation between blockchains to solve the problem of fraud on the consortium chain. Instead, the Bitcoin payment channel provides end-to-end security to users, which is logically separated from the consortium chain. Our approach is simple, but has the flexibility to connect to various consortium chains without being bound by a specific specification. Niji can also be easily combined with existing solutions to improve its abilities.

\subsection{Feasibility}
We evaluated the feasibility of the Niji protocol between the Bitcoin test-net and an EVM-based consortium chain built using the Ethereum client. The prototype was implemented using the bitcoin-core client and go-ethereum client, which supports clique proof-of-authority \cite{clique} as a consensus algorithm for building consortium chains \cite{cons}. We placed four nodes with authority to generate blocks in the Ethereum-based consortium chain and changed the period of proof-of-authority consensus from a default 15 seconds to 1 second to get those blocks mined faster. The Bitcoin client is only used to broadcast Bitcoin transactions between setup and settlement. We confirmed that all Bitcoin transactions in the Niji protocol were accepted into Bitcoin test network.
In this evaluation, we focused on the performance related to consortium chain transactions. Additionally, to clarify only the payment performance of Niji, we excluded measurements related to service provision in the service contract.

\subsubsection{Computational cost}
\begin{table}[tbp]

\renewcommand{\arraystretch}{1.2}

\caption{Gas Cost of Computational Work}
\label{gas_cost}
\centering

\begin{tabular}{lrr}
 \toprule Operation & \multicolumn{2}{c}{Gas cost (gas)} \\
 \cmidrule(lr){2-3}
  & uni-directional & bi-directional \\
 \midrule
 Deploy bridging contract & 1,785,044 & 2,562,757\\
 Set deposit information & 245,922 & 245,988\\
 Set transaction template & 285,183 & 431,100\\
 Update payment & 455,355 & 468,389\\
 Replace transaction template & - & 431,164\\
 Disclose pre-image & - & 53,687\\
 Close channel & 46,626 & 46,648\\
 \bottomrule
\end{tabular}
\end{table}

The bridging contract coded by Solidity, Ethereum’s most popular high-level language, exposes several API calls to receive transactions. These operations consume {\it gas}, the internal pricing used for executing operations in the EVM. Each computation of the contract codes has an associated cost in gas, protecting the blockchain network from denial-of-service attacks with infinite loops and encouraging efficiency in the code. The gas system is also useful on the consortium chain in terms of forcibly terminating harmful code execution, even if the gas itself has no value (i.e., users need not pay for gas with crypto-currency) unlike in the public Ethereum blockchain. In fact, Quorum, the EVM-based consortium chain from JP Morgan, removes pricing of gas, although the gas system itself remains \cite{quor}.

The transaction gas costs of operations through the bridging contract are listed in Table III. These values come within the usual range found in a public Ethereum block, which has a gas limit of approximately 8.0 million (in June 2018). The bi-directional channel tends to require higher gas cost than the uni-directional one for the management of the cancellation procedure. When constructing the consortium chain, the gas limit in each block and the amount of gas included in a transaction should be adjusted with reference to these results. In actual applications, since a payment calls the function of the service contract automatically, the corresponding gas cost for the service is added to the gas value of updating payment shown in Table III.

\subsubsection{Timing analysis}
\begin{table}[tbp]

\renewcommand{\arraystretch}{1.2}

\caption{Time Analysis of Operations in Payment Phase}
\label{timing}
\centering

\begin{tabular}{lcccc}
 \toprule Task & Mean & Maximum & Minimum & \shortstack{Std. \\ Deviation}\\
 \midrule
 Get transaction template & 116 & 454 & 101 & 18.0 \\
 Update payment & 1981 & 2664 & 1112 & 252 \\
 Total & 2097 & 2778 & 1218 & 253 \\ 
\end{tabular}
\end{table}
Table IV outlines the timing analysis measurements for the payment phase in Niji with bi-directional channel. All times are in milliseconds. We executed 1000 experimental runs and computed latency statistics including the average (mean), maximum, minimum, and standard deviation. All measurements were performed on a MacBook Pro running OS X 10.13.4 equipped with four cores, 2.3GHz Intel Core i5, and 16 GB memory. One payment operation was divided into two tasks: getting the transaction template and updating payment. ``Total" is the end-to-end response time as defined in one payment. We used Node.js as the measurement environment of the Niji protocol to communicate with go-ethereum and bitcoin-core. Each operation included not only the processes in Node.js but also those in the EVM, such as verifying a signature and parsing a transaction.

In the process of getting a transaction template, the user never broadcasts transactions. Since data is only retrieved from the blockchain, the process runs in a steady rate. On the other hand, payment update, which has time ranges from $1121$ ms to $2664$ ms, is clearly dominated by the consensus time, i.e., the time until the registered data is stored in a new block. In this experiment, we set the parameter of the consensus time to 1 second; therefore, $2094-1000 = 1094$ ms is the overhead for each payment. Although the payment can be updated many times as long as the channel is open, a certain latency as above is required for each payment. Therefore, services using the Niji payment protocol must be designed with this constraint in mind.

Moreover, although scalability is an aspect of performance, it entirely depends on the throughput of the consortium chain itself. The Niji protocol certainly contributes to an improvement of the Bitcoin scalability, but the choice of consortium chain platform may have a great effect on this aspect. In the next section, we discuss possible extensions to Niji, including ones aimed at scalability.

\section{Discussion}
Here, we compare Niji with other cross-chain protocols and discuss possible extensions of the Niji protocol to a wide range of future applications. These extensions could remain within the basic Niji concept or be combined with existing solutions to improve its abilities.

\subsection{Comparison with related work}
Several cross-chain protocols have been proposed for improving scalability and interoperability. These approaches can be classified into two types: utilizing intermediaries like a federation or decentralized networks and directly connecting blockchains with the cooperation of participates. The former way has a premise that the participants place trust in the intermediaries. Federated sidechain\cite{fed} relies on a federation for honest activity and the federation controls multi-signature locks to transfer coins between Bitcoin and the sidechain. As more decentralized approaches, Cosmos\cite{cosmos} and Polkadot\cite{polka} have an inter-blockchain to relay data between blockchains, respectively referred to as Cosmos hub and Relay chain. The inter-blockchain is operated by incentivized nodes, but the fact that participants rely on validators is similar to the federation. Although these approaches certainly need to place trust in intermediaries, their participants are less burdened, as the intermediaries are responsible for the connections and monitoring costs.

On the other hand, in the latter way, i.e., directly connecting blockchains with the cooperation of participates, individual participants themselves need to take precautions to prevent fraud. Atomic swaps \cite{atomic} (or atomic cross-chain trading) is the exchange of one cryptocurrency with another cryptocurrency between two parties, without the need to trust a third party. The tokens are directly traded between users in a trustless atomic manner, but both users must continuously monitor the blockchains to create and broadcast the transaction synchronously together, which places a burden on them. Plasma is a powerful framework that makes smart-contract execution scalable by associating block creations in the child blockchain with the root Ethereum blockchain. This approach also requires periodic commitments, i.e., submission of block headers to the root blockchain. In addition, to prevent fraud, it is necessary for users to constantly monitor both parent and child chains; in general, the latter way tends to burden individual participants.

Although Niji is classified as the latter sort of method, it has a different aspect from other related work. The key idea is that there is no value transfer of bitcoins to tokens of another blockchain. In most of the related work, by bounding funds in a particular contract, you can swap them with another token that holds the same value as the bounded funds in another blockchain. If another blockchain (i.e., consortium chain) fails to work properly due to misconduct or fraud, funds will be not protected at all, unless there is an incentivized and massive mechanism protects the funds like in Plasma. By contrast, in Niji, value transfer of bitcoins does not occur between blockchains. Instead, the value of bitcoins can be used directly and circulated on the consortium chain. Niji has no tight cooperation between blockchains, but it achieves a simple and flexible connection to another blockchain, which has certain security based on the payment channel. Our protocol design reduces the monitoring costs and its concept eliminates the risk of price fluctuations in tokens. As an analogy, the U.S. dollar can be used as legal tender in some countries where either no local currency is issued or both the local currency and US dollar coexist. ``Dollarization" stabilizes the value of the local currency, which leads to stabilization of prices and the economy of the whole country; similarly, ``Bitcoinization" in consortium chains that built by startups or enterprises would bring about stable development and growth for their services.

\subsection{Niji in combination with other platforms}
The Niji implementation described in this paper uses an EVM-based consortium chain platform, and Bitcoin is the only cryptocurrency supported. Of course, the Niji protocol can easily be extended to other platforms, e.g., a combination of Litecoin and Hyperledger Fabric if their requirements can be satisfied. The requirement of available currency is to have a scripting function similar to Bitcoin, that is, the currency has a multi-signature address and the time-lock and hash-lock functionalities. On the other hand, the choice of consortium chain platform has a direct impact on the scalability of Niji payments. Hyperledger Fabric has the ability to handle Turing-complete smart contracts similarly to EVM-based platforms and as well provides high throughput performance \cite{perm}. Since the Niji protocol is designed to run even in restricted environments like EVM, it may work on many consortium platforms including Hyperledger. This would enable appropriate platforms to be selected according to the use case.

\subsection{Possible improvement with eltoo channel}
Decker, Russell, and Osuntokun have proposed the eltoo protocol \cite{eltoo} as another approach to realizing a simpler and more efficient bi-directional channel. Unlike the Lightning Network channel, this novel channel does not require penalty branches. It thus reduces monitoring costs and overcomes the problem of asymmetry of information that endpoints have. The protocol works by overriding the previous state and forces old transactions to be unavailable. eltoo introduces the concept of state numbers, similar to sequence numbers in the original Nakamoto implementation of Bitcoin. While the original sequence numbers of the Nakamoto implementation were not enforceable, eltoo's state numbers enforce replacement of transactions by ensuring that a later state can resume any of the previous states until the last settlement transaction is confirmed. In order to achieve the replacement mechanism, it is necessary to introduce a new sighash flag, SIGHASH\_NOINPUT, which selectively marks transactions that can be tied to previous transactions.

The eltoo protocol can be applied to Niji, because it does not require interactive exchange of something to prevent counterparty fraud when updating a payment. The main benefit of using eltoo would be that it simplifies the cancellation of payment procedure. In Niji, the state number is managed by a smart contract, and the latest state always has priority regardless of whether the money sent to the service provider is increased or a payment is cancelled. The cancellation procedure described in section IV-B requires four steps, but with eltoo, it would be shortened to two steps: the user requests to cancel a payment and then the service provider agrees to it by submitting its signature. In addition, it is also possible for the service provider to refund the user by unilaterally submitting its signature.

In order to enable eltoo protocol on Niji, we need to modify the template transactions slightly. The eltoo channel requires two types of transaction, one for update and one for settlement, and a settlement transaction always requires a new public key pair. These features might entail modifications to the verification process in the bridging contract. More importantly, it is necessary that the eltoo protocol itself become available in the Bitcoin protocol; that is, the new features that compose eltoo, state numbers and SIGHASH\_NOINPUT, must be made available in the Bitcoin blockchain infrastructure in the future.

\section{Conclusion}
We presented Niji, a cross-chain protocol for atomic and secure Bitcoin payment on a consortium blockchain. This protocol provides a means of payment into a consortium chain using the Bitcoin payment channel. The process autonomously runs from payment to service provision with no need for trusted third parties, and with no transfer of bitcoins to other tokens. Our first implementation is designed to work on an EVM-based consortium chain, and our experiments showed the practical feasibility of the protocol. We discussed the security of Niji as well as future extensions that could make use of existing platforms to improve the functionality and scalability of the protocol. These extensions could potentially make payments on the consortium chain more flexible and efficient for a wide range of applications.

\begin{table}[!b]
\renewcommand{\arraystretch}{1.0}
\caption{Structure of Transaction Template in Segwit Use}
\label{timing}
\centering
\begin{tabular}{llll}
 \toprule Field &  & Value (example) \\
 \midrule
 Version &  & 01000000 \\
 Marker &  & 00 \\
 Flag &  & 01 \\
 Input count &  & 01 \\
 \cmidrule(l){1-2}\cmidrule(lr){3-3}
 Input {[}0{]} & Previous output & \textless{}funding tx hash\textgreater{} \\
  & Index & \textless{}index of previous output\textgreater{} \\
  & Script & OP\_0 \textless{}hash of witnessScript\textgreater{} \\
  & Sequence &  FFFFFFFF \\
 \cmidrule(l){1-2}\cmidrule(lr){3-3}
 Output count &  & 02 \\
 \cmidrule(l){1-2}\cmidrule(lr){3-3}
 Output{[}0{]} & Value & {\bf nil} \\
  & Script & OP\_0 \\
  & &\textless{}service provider's pubkey hash\textgreater{}\\
 Output{[}1{]} & Value & {\bf nil} \\
 (change given & Script & OP\_0 \\
 back to user)& & \textless{}user's pubkey hash\textgreater{} \\
 \cmidrule(l){1-2}\cmidrule(lr){3-3}
 Witness count & & {\bf nil} \\
 Witness & & only \textless{}witnessScript for previous\\
 & & output\textgreater{} {\bf (no signatures)} \\
 Locktime &  & 00000000 \\
 \bottomrule
\end{tabular}
\end{table}

\section*{Appendix}

\subsection{Segwit templates}
The transaction template must be modified if the Segwit transaction is used. The modification introduces two types of template: a Segwit transaction itself and a serialization format for SignatureHash algorithm specified in BIP143\cite{bip143}, respectively shown in Table V and Table VI. The format in Table VI is used as hash pre-image corresponding to the signature included in the Segwit transaction, but its template is missing $hashOutputs$ field that requires sha256 double hash of the serialization of output amount with scriptPubKey. When updating a payment, the signature submitted by the user can be verified by combing these templates.

\begin{table}[tbp]
\renewcommand{\arraystretch}{1.0}
\caption{Format Template For Segwit Signature Verification}
\label{timing}
\centering
\begin{tabular}{llll}
 \toprule Field &  & Value (example)\\
 \midrule
 Version &  & 01000000 \\
 hashPrevouts &  & \textless{}32-byte hash of all outpoints\textgreater{}\\
 hashSequence  &  & \textless{}32-byte hash of sequence of all inputs\textgreater{} \\
 Outpoint  &  & \textless{}funding tx hash and index\textgreater{} \\
 scriptCode &  & \textless{}witnessScript\textgreater{} \\
 Value of previous output &  &  \textless{}deposit amount of funding tx\textgreater{}\\
 Sequence of input  &  & FFFFFFFF \\
 hashOutputs & & {\bf nil} \\
 Locktime & & 00000000\\
 Hash Type &  & 01000000\\
 \bottomrule
\end{tabular}
\end{table}

\subsection{Bitcoin signature verification on EVM}
The verification of Bitcoin's signature on EVM depends on Solidity, a language of smart contracts in Ethereum. The Solidity specification defines a precompiled contract $EcdsaRecover$ as follows:
\begin{lstlisting}[linewidth=\columnwidth,breaklines=true]
ecrecover(bytes32 hash, uint8 v, bytes32 r, bytes32 s) returns (address)
\end{lstlisting}
where the $ecrecover$ function recovers an Ethereum address associated with the public key from the elliptic curve signature. As described in section III-B, to make a comparison with the recovered Ethereum address, a Bitcoin public key corresponding to the signature needs to be converted into an Ethereum-style address. According to the Ethereum yellow paper \cite{pcon}, an Ethereum address $A(pr)$ (a 160-bit value) for a given Bitcoin private key $pr$ is computed as follows:
\begin{equation}
A(pr) = B_{96..255}(KEC(BITCOINPUBKEY(pr)))
\end{equation}

where $A(pr)$ is defined as the right-most 160 bits of the Keccak hash of the corresponding Bitcoin public key; the Bitcoin public key virtually is converted into an Ethereum address.

Using the $ecrecover$ function and virtual Ethereum address, the $verify$ function in a smart contract can be implemented as follows:

\begin{lstlisting}[linewidth=\columnwidth,breaklines=true]
function verify(bytes32 modTx, uint8 v, bytes32 r, bytes32 s, address convertedAddr) public pure returns(bool) {

    bytes32 hash = sha256(sha256(modTx));
    address Addr = ecrecover(hash, v, r, s);

    if(Addr == convertedAddr) return true;
    else return false;
}
\end{lstlisting}

The $verify$ function receives an unsigned transaction $modTx$, signature parameters $(v,r,s)$, and $convertedAddr$ from the Bitcoin public key, and it outputs the results of the verification. In the Niji protocol, this verification process is included in the bridging contract.

\end{document}